\documentstyle[11pt,newpasp]{article}

\begin{document}

\title{Intracluster Planetary Nebulae in the Virgo cluster}
\author{K.C. Freeman$^1$}
\author{M. Arnaboldi$^2$, M. Capaccioli$^2$, R. Ciardullo$^3$, 
J. Feldmeier$^3$, H. Ford$^4$, O. Gerhard$^5$, R. Kudritzki$^6$, 
G. Jacoby$^7$, R.H. M\'endez$^6$, R. Sharples$^8$}

\affil{$^1$Research School of Astronomy \& Astrophysics, Mt. Stromlo 
Observatory, Canberra, Australia\\
$^2$Osservatorio Astronomico di Capodimonte, Naples, Italy\\
$^3$Penn State University, University Park, U.S.A.\\
$^4$John Hopkins University, Baltimore, U.S.A.\\
$^5$Institute for Astronomy, University of Basel, Binningen, Switzerland\\
$^6$Munich University Observatory, Munich, Germany\\
$^7$Kitt Peak National Observatory, Tucson, U.S.A.\\
$^8$University of Durham, Durham, U.K.}

\begin{abstract}
We briefly describe the properties of the confirmed spectroscopic sample of 
the intracluster planetary nebulae recently discovered in the Virgo cluster. 
We find 23 ``{\em bonafide}'' intracluster planetary nebulae, 
and 8 high redshift ($z\sim 3.1$) Ly$\alpha$ emitters identified 
by their broad asymmetric emission line.
\end{abstract}


\section{Introduction}
Planetary nebulae (PNe) are excellent tracers of very diffuse stellar 
populations:
e.g. outer region of early type galaxies (Ciardullo et al. 1993, 
Hui et al. 1995, Arnaboldi et al. 1998), and of the intracluster stellar 
medium in cluster of galaxies. With 4 meter telescope one can detect 
individual PN and measure their velocities out to 20 Mpc.

How are extragalactic PNe identified? Images of a cluster field
or in the outer region of galaxies are acquired with a narrow filter whose 
central wavelengths coincide with the redshifted 5007 \AA\ [OIII] emission 
line of a PN belonging to the system. 
The bandwidth $\Delta \lambda$ is taken to be $ 6 \times$ the velocity
dispersion of the system under study; this width corresponds to 
$\Delta \lambda = 40$ \AA\ for large ellipticals and to 
$\Delta \lambda = 65$ \AA\ for the Virgo cluster field. 
Then an off-band image is taken in the adjacent continuum of the
[OIII] 5007 \AA\ line with a bandwidth of $200 - 1000$ \AA. 
With 4 meter telescopes on observing sites with $\sim 1''$ seeing, a 5 hrs 
on-band image gives detections complete to 
$F_{5007} = 5\times 10^{-17}$ ergs cm$^{-2}$ s$^{-1}$ or m$_{5007} = 27.0$. 
This covers the first magnitude range of the PNLF at the Virgo cluster 
distance.\\
Most work on intracluster PNe (IPNe) has been done on the Virgo cluster
(distance $\simeq 15$ Mpc). The first spectroscopic confirmation
was from Arnaboldi et al. (1996) who detected 3 PNe with 
radial velocities $\ge 1300$ kms$^{-1}$ in the M86 field: since 
M86 has a peculiar negative redshift at $v_{sys} = -227$ kms$^{-1}$,
these PNe must belong to the Virgo cluster.
M\'endez et al. (1997)  surveyed a blank field in the Virgo cluster core 
with the 4 meter WHT and found 11 [OIII] emission line candidates in a 50 
arcmin$^2$ field. Feldmeier et al. (1998) surveyed several fields in the 
Virgo cluster and discovered 150 [OIII] emission line candidates in the 750 
arcmin$^2$ total surveyed area.
 
If these [OIII] emission line candidates are all associated to 
the 5007 [OIII] emission from PNe, one can estimate the light associated
with the stellar parent population of the intracluster PNe. 
From the above surveys, the parent population contributes about 50\%
of the smoothed-out surface brightness in the cluster core.
Evidence for an evolved stellar population in the Virgo cluster 
was also found by direct imaging of red giant stars with the HST. 
Ferguson et al. (1998) found 600 probable red giants in a deep WFPC2 
I-band image in Virgo cluster core.

\section{Observations}
The [OIII] emission line candidates from the narrow band imaging surveys
of M\'endez et al. (1997) and Feldmeier et al. (1998) were
observed with the 2 degree field fiber spectrograph (2dF) at the 
Anglo Australian 4 meter telescope. The spectra for the [OIII] emission line
candidates were obtained with a spectral resolution 
R=2000 and a total integration time of 5 hours on the 15th of March 1999.
Given a target sample,
the 2dF software chooses randomly out of the sample the maximum number of 
candidate positions to which the fibers are allocated. Some fibers
were also placed on PNe candidates in the outer regions of M87,
from Ciardullo et al. (1998).
\begin{figure*}\label{starburst}
\includegraphics{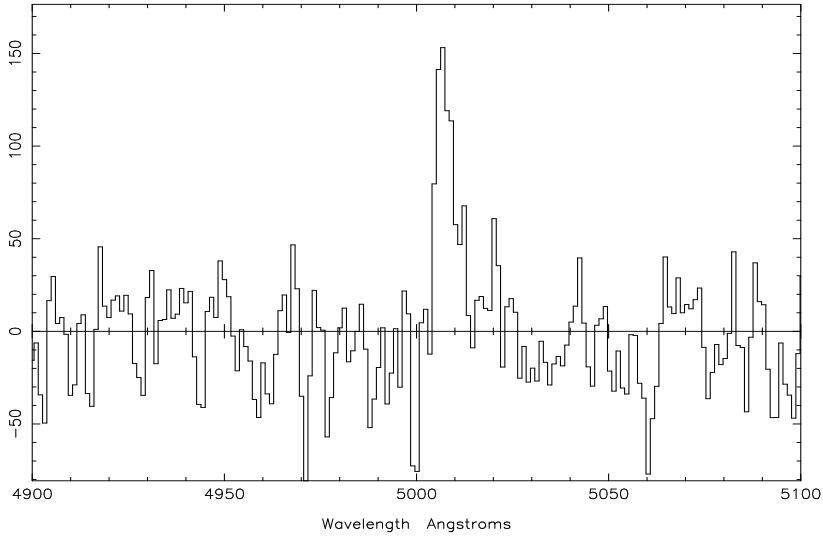}
\vspace{7cm}
\caption{2dF spectrum of a probable starburst galaxy at z=3.14.}
\end{figure*}
We got 47 detections out the 110 fibers allocated. All detections are
for [OIII] emission line candidates whose m$_{5007} < 27.0$.
\begin{itemize}
\item 23 turn out to be real intracluster PNe of the Virgo cluster.
The [OIII] 4959 \AA\ line was detected together with the 5007 \AA\ line.

\item 16 PNe were detected in the outer regions of M87 ($\bar{R} = 24$ kpc).

\item 8 spectra showed a broad asymmetric single line: 
probably Ly$\alpha$ from starburst at $z \simeq 3.1$, with
an equivalent width $W_\lambda (\mbox{Ly}\alpha) > 150$ \AA.
One of these spectra is shown in figure~1.
\end{itemize}

\subsection{Check on reality of IPN detections} 
Because of the limited S/N ratio, we did not see both of the
[OIII] lines in all of the 23 likely IPN detections.
The PN [OIII] 4959/5007 \AA\ emission lines have a fixed
flux ratio: $I_{5007} = 3 \times I_{4959}$. We can use this as
a way of checking on the reality of the 23 IPN spectra. For each of the
23 spectra 1) we normalise the peak $5007$\AA\ intensity to
100, 2) shift spectrum to zero velocity, 3) then add them. 
If all are real IPNe, i.e. not noise or redshifted galaxies with high
$W_\lambda$, then one should expect 
$
\frac{\mbox{I}_{5007}}{\mbox{I}_{4959}} \sim 3.0  
$
in summed spectra. The observed ratio for the 23 summed spectra is 
$3.2 \pm 0.1$, so real fraction is $0.94 \pm 0.03$. The summed spectrum is
shown in figure~2.

\begin{figure*}\label{sum}
\includegraphics{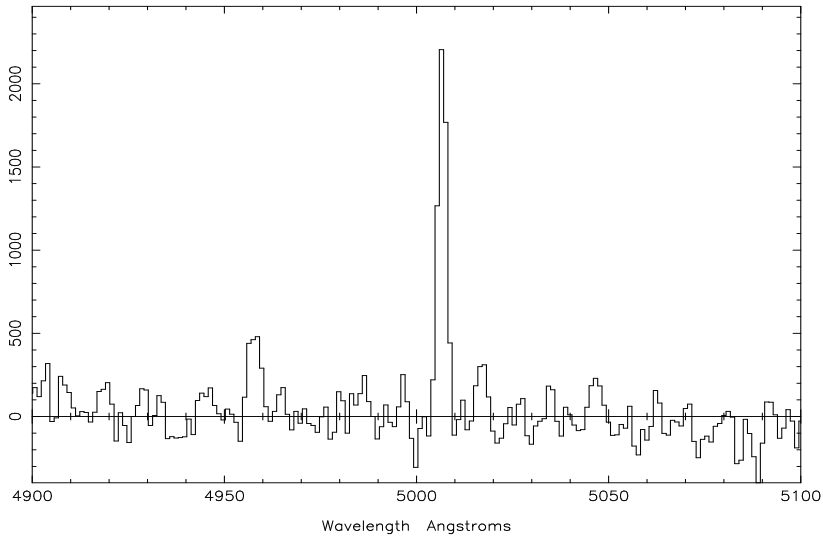}
\vspace{7cm}
\caption{Sum of 23 IPN spectra in Virgo. The real fraction of 
IPNe is $0.94\pm 0.03$.}
\end{figure*}
The spectroscopic survey of [OIII] emission line candidates in
the Virgo cluster has a success rate of 55\%$\pm 17$ for
candidates in the M87 field, and 38\%$\pm 8$ for the intracluster
[OIII] emission line candidates with m$_{5007} < 27.0$.
The failure to detect spectra of our candidates can be due to either to
1) candidates being not emission line objects, or 2) technical problems, e.g.
differential refraction over long exposures or astrometry. 

\section{Discussion}

{\em Velocity distribution of IPNe, M87 PNe and Ly$\alpha$ objects --}
We studied the properties of the radial velocity distribution
for the  spectroscopically confirmed objects.
For the confirmed 23 IPNe , the average velocity is
$\bar{v} = 1193 \pm 156$ kms$^{-1}$ and the velocity dispersion
of the sample is $\sigma = 752 \pm 110$ kms$^{-1}$.  These values
are in complete agreement with those determined from 
the galaxy radial velocities in the Virgo cluster, i.e.
$\bar{v}_{cluster} = 1100$ kms$^{-1}$ and $\sigma_{cluster} = 800$ kms$^{-1}$.
The 16 M87 PNe in our spectroscopic sample have an average
$\bar{v} = 1413 \pm 74$ kms$^{-1}$ and $\sigma = 287 \pm 51$ kms$^{-1}$.
These values agree with those detemined for the globular cluster system in M87,
whose average velocity and sigma are $\bar{v} = 1280$ kms$^{-1}$
and $\sigma = 340 \pm 30$ kms$^{-1}$.
M87 globular clusters show $\simeq 100$ km s$^{-1}$ rotation: 
our PNe in M87 are mostly on the high velocity side of M87.
On the other hand, the Ly$\alpha$ objects turn out to be uniformly 
distributed  through out the velocity range allowed by the narrow filter 
bandpass used to select the [OIII] emission line candidates. \\
{\em On the contamination by high-z galaxies --}
From the narrow band imaging surveys of M\'endez et al. (1997) and
Feldmeier et al. (1998), the m$_{5007}$ magnitudes of the 
spectroscopically confirmed candidates are all brighter than 27.0, which
implies a total flux in the [OIII] 5007 line of 
$F_{5007} > 5 \times 10^{-17}$ ergs cm$^{-2}$ s$^{-1}$ and 
$W_\lambda \ge 150$ \AA, i.e. no detection in the off-band image. 
From narrow imaging surveys done on control fields (Arnaboldi, Feldmeier,
1999, in preparation)  one would expect $95 \pm 27$ objects per deg$^2$ 
to the limiting magnitude $m_{5007} = 27.0$.
In our sample, we expect $14 \pm 4$ of which we would have detected $6 \pm 2$;
we detected 8 of such objects.

\section{Conclusions}

The distribution of the IPNe radial velocities in the Virgo cluster has 
similar $\bar{v}, \, \sigma$ to the cluster galaxies. 
The IPNe sample is still too small to see spatial and velocity structure - much
larger survey is in progress.
Of the emission-line candidates which we detected (all with m$_{5007} < 27.0$),
about 25\% of the candidates indentified in narrow 
band survey are high-z Ly$\alpha$ emitters. The other 75\% of intracluster
candidates are almost all IPNe.

\acknowledgments
We wish to thank the Anglo Australian telescope and the 2dF team
for their support during our observing run.


\begin{references}
\reference Arnaboldi, M., Freeman, K.C., et al. 1996, \apj, 472, 145
\reference Arnaboldi, M., Freeman, K.C., et al. 1998, \apj, 507, 759
\reference Ciardullo, R., Jacoby, G.H., Dejonge, H.B., 1993, \apj, 414, 454
\reference Ciardullo, R., Jacoby, G.H., et al. 1998, \apj, 492, 62
\reference Feldmeier, J.J., Ciardullo, R., Jacoby, G.H. 1998, \apj, 503, 109
\reference Hui, X., Ford, H., et al. 1995, \apj, 449, 592
\reference M\'endez, R. H., Guerrero, M. A., et al. 1997, \apjl, 491, L23
\end{references}
\end{document}